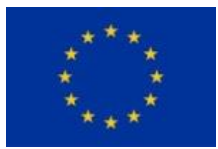
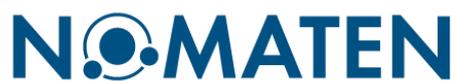

This work was carried out in whole or in part within the framework of the NOMATEN Centre of Excellence, supported from the European Union Horizon 2020 research and innovation program (Grant Agreement No. 857470) and from the European Regional Development Fund via the Foundation for Polish Science International Research Agenda PLUS program (Grant No. MAB PLUS/2018/8), and the Ministry of Science and Higher Education's initiative "Support for the Activities of Centers of Excellence Established in Poland under the Horizon 2020 Program" (agreement no. MEiN/2023/DIR/3795).

The version of record of this article, first published in Spectrochimica Acta Part A: Molecular and Biomolecular Spectroscopy, Volume 321, 15 November 2024, 124680, is available online at Publisher's website: https://dx.doi.org/10.1016/j.saa.2024.124680

*Study of amorphous alumina coatings for next-generation nuclear reactors: high-temperature in-situ and post-mortem Raman spectroscopy and X-ray diffraction*

Magdalena Gawęda[1]*, Piotr Jeleń[2], Agata Zaborowska[1], Ryszard Diduszko[1,3], Łukasz Kurpaska[1]

[1] NOMATEN CoE, NOMATEN MAB, National Centre for Nuclear Research, A. Soltana 7 Str., 05-400 Otwock-Świerk, Poland
[2] Faculty of Materials Science and Ceramics, AGH University of Kraków, A. Mickiewicza 30 Av, 30-059 Kraków, Poland
[3] Institute of Microelecotronic and Fotonics, Łukasiewicz Research Network, Wólczyńska 133 Str., 01-919 Warsaw, Poland

*[Magdalena.Gaweda@ncbj.gov.pl](mailto:Magdalena.Gaweda@ncbj.gov.pl)*

**Abstract**

The present work focuses on the investigation of the thermal stability and structural integrity of amorphous alumina coatings intended for use as protective coatings on cladding tubes in Generation IV nuclear reactors, specifically in the Lead-cooled Fast Reactor (LFR) type. High-temperature Raman spectroscopy and high-temperature X-ray diffraction analyses were carried out up to 1050 °C on a 5 μm coating deposited by the pulsed laser deposition (PLD) technique on a 316L steel substrate. The experiments involved the *in-situ* examination of structural changes in the material under increasing temperature, along with *ex-situ* Raman imaging of the surface and cross-section of the coating after thermal treatments of different lengths. As it was expected, the presence of α-alumina was detected with the addition of other polymorphs, γ- and θ-$Al_2O_3$, found in the material after longer high-temperature exposure. The use of two structural analysis methods and two lasers excitation wavelengths with Raman spectroscopy allowed us to detect all the mentioned phases despite different mode activity. Alumina analysis was based on the emission spectra, while substrate oxidation products were identified through the structural bands. The experiments depicted a dependence of the phase composition of oxidation products and alumina's degree of crystallization on the length of the treatment. Nevertheless, the observed structural changes did not occur rapidly, and the coating's integrity remained intact. Moreover, oxidation signs occurred locally at temperatures exceeding the LFR reactor's working temperature, confirming the material's great potential as a protective coating in the operational conditions of LFR nuclear reactors.

Keywords: HT-Raman; HT-XRD; Raman imaging; amorphous alumina coating; $Al_2O_3$; generation IV nuclear reactor

## 1. Introduction

The severe climate crisis drives demand for new low- and non-emissive energy sources, with emphasis being placed on nuclear power. Despite the high performance of current technologies, novel concepts are being developed to surpass contemporary designs in terms of sustainability, reliability, and safety. One of the emerging concepts among Generation IV nuclear reactors is the Lead-cooled Fast Reactor (LFR) [1].

The sustainability aspect of the LFR is ensured by its use of a fast-neutron spectrum, a closed fuel cycle, and the ability to utilize depleted uranium or thorium fuel matrices, as well as actinides spent fuel from the light-water reactor (LWR). The safety of the LFR is ensured by the unique coolant: liquid lead, which has low neutron moderation and absorption capacity, a very high boiling temperature (1749 °C), operation pressure close to atmospheric, chemical inertness to air and water, and fission product retention. These features make the LFR concept economically competitive, but this will only be the case if reliability is ensured, which depends heavily on the construction materials.

Since the operational lifetime of a nuclear facility is long (in this case, over 40 years), the materials used in its construction must be able to withstand high irradiation levels (up to 200 dpa), elevated temperatures (up to 600 °C), and contact with corrosive coolant (liquid lead or lead-bismuth eutectic (LBE)) [1], creating a challenging work environment that elevates the material requirements. For this reason, the LFR concept design requires materials with high stability at high temperatures, excellent corrosion resistance, sufficient mechanical properties, and good radiation tolerance.

Available austenitic stainless steels (e.g., 316L, 15-15 Ti) or ferritic/martensitic steels (e.g., T91, HT9) are unsuitable for use in these types of coolants at a temperature above 500 °C due to the deterioration of the material integrity caused by corrosion and lead infiltration [2, 3]. Over the past decade, several solutions have been proposed, including the development of new iron-based Al-containing alloys that form an auto-forming protective alumina scale (e.g., AFA) [4] or the application of protective amorphous ceramic coatings, i.e., SiOC [5], or alumina [6], which is the focus of this study.

In this work, we investigate the proposed solution of amorphous alumina coatings deposited with pulse laser deposition (PLD) on metallic substrates [7]. These coatings possess outstanding functional properties that are crucial for their application in LFR. This is attributed to their strong alumina/steel interfacial bonding, compatibility with the substrate's mechanical properties, and moderate hardness. They also exhibit satisfactory adhesion and exceptional wear resistance [6]. Moreover, they demonstrate good resistance to lead-based coolants [8, 9].

Previous research extensively reported on their behaviour after irradiation, documenting changes in structure, microstructure, and mechanical properties [10, 11]. It has been proven that the alumina coatings remain amorphous under gold ion irradiation with no evidence of void formation, phase segregation, or crystallisation. Regarding mechanical parameters, the slight evolution observed contributes to even better adaptation to the substrate's parameters ensuring compatibility of the material as the coating/metal composite [10].

However, structural degradation under elevated temperatures, another concern in the hostile LFR reactor environment, is inevitable and requires careful consideration. Particularly valuable are *in situ* observations of the impact of elevated temperatures on the material's behaviour, complemented by the post-mortem studies of the system when cooled down to room temperature and, hence, the analysis itself is facilitated. Unfortunately, the literature currently lacks *in-situ* high-temperature structural analysis data on the alumina amorphous coatings.

Observing material with high-temperature X-ray diffraction enables the tracking of only long-range structural changes. In contrast, Raman spectroscopy is more sensitive to short-range order and can serve as the complementary analytical method. Through this technique, phase transformation can be observed, and subtle spectral changes may reveal physical phenomena associated with thermal expansion, variations in the population of vibrational energy levels, bond length, and vibrational force constant with increasing temperature.

The Raman spectroscopy analysis conducted in this study focused on identifying the various polymorphs of alumina, including $\alpha$-$Al_2O_3$, $\gamma$-$Al_2O_3$, and $\theta$-$Al_2O_3$. The Raman spectrum of $\alpha$-$Al_2O_3$ is well-characterized, with distinct bands associated with its hexagonal crystal structure at approximately 378 $cm^{-1}$, 417 $cm^{-1}$, and 450 $cm^{-1}$, corresponding to the vibrational modes of the corundum structure. The prominent band around 417 $cm^{-1}$ corresponds to the symmetric stretching mode of the aluminum-oxygen bond (ν1 mode), as well as other bands in the regions corresponding to the $A_{1g}$ and $E_g$ vibrational modes [12, 13]. γ-$Al_2O_3$, a metastable phase, shows a different Raman signals. Its bands are often broader and less defined due to its defect-rich, spinel-like structure. Key Raman bands for γ-$Al_2O_3$ can be found in regions indicative of tetrahedral and octahedral cation vibrations around 370 cm$^{-1}$, 490 cm$^{-1}$, and 720 cm$^{-1}$ [14]. Furthermore, θ-$Al_2O_3$ is an intermediate phase between γ- and α-$Al_2O$, and its Raman spectrum reflects this transition. It exhibits characteristic bands that are distinct from those of both γ- and α-phases near 380 cm$^{-1}$ and 660 cm$^{-1}$, associated with the orthorhombic structure of θ-$Al_2O_3$, aiding in its identification during phase transformations [15, 16].

This work presents alternative way of analysis of presence of alumina polymorphs. Structural bands for $Al_2O_3$ are relatively weak compared to bands with origin in steel substrate corrosion products and the emission bands which may be induced by the red lase line. Despite Raman spectroscopy is not specifically designed for this purpose, emission bands may be observed along with the Stokes scattering corresponding to the classical Raman spectra [17] which ensures accurate phase identification. This approach allows for a more reliable detection of these phases despite the lower intensity of their structural Raman signals. α-$Al_2O_3$ exhibits characteristic emission bands in the visible range, typically around 400-700 nm, arising from defect-related transitions within the crystal lattice [13, 18]. The emission bands of γ-$Al_2O_3$ and θ-$Al_2O_3$ are less studied compared to α, but they are believed to arise from similar defect-related transitions, albeit with different energy levels due to variations in the local environment and crystal structure [16, 19].

Unfortunately, several effects, such as fluorescence, may hinder Raman spectrum, not only in high-temperature measurements. Despite these limitations, our XRD and Raman measurements conditions allowed us to observe initiation of crystallisation of α-, γ- and θ-$Al_2O_3$ in the analysed amorphous alumina coating, along with the appearance of steel substrate oxidation products. The presented research provides a basis for evaluating the longevity of the material when applied in the LFR reactors.

## *2. Methodology*

The 5 μm alumina coatings under investigation were deposited on 0.9 mm austenitic 316L stainless steel using pulsed laser deposition (PLD) at room temperature [20]. Obtained samples were cut to appropriate sizes for further measurements with wire electrical discharge machining (WEDM).

High-temperature X-ray diffraction (HT-XRD) was conducted using a Rigaku SmartLab 3kW diffractometer and the dedicated high-temperature chamber in Bragg-Brentano geometry with 0.02° 2θ step. The same experimental conditions were used at room temperature (35 °C) and during the *in-situ* high-temperature tests, which were conducted within a temperature range of 150-1050 °C with a 100 °C step. A heating rate of 30 °C/min was employed, and each measurement was held at the target temperature for 40 min.

The cross-section of the sample was prepared by first Au sputtering (~10 nm layer), followed by nickel-plating (~1 ÷ 2 μm) using a Watts bath (a mixture of nickel sulphates and chlorides). The sample was then embedding in a carbon-based conductive resin. The final step involved metallographic preparation, including water grinding with 220 Piano and polishing with diamond pastes using Allegro-9 μm /Mol-3 μm / Nap-1 μm grinding and polishing systems.

High-temperature Raman spectroscopy was performed using a WITec alpha 300M+ microspectrometer and Linkam TS1500 chamber. To prevent interferences and band overlap with the analysed alumina coating, the standard sapphire top window of the chamber was replaced with a silica window. The first tests were conducted within a temperature range of 50-1050 °C with a 50 °C step, 5 °C/min heating rate, and 5 min temperature step maintenance time. The second adopted methodology included analogical parameters with the exception of maintenance time, which was elongated to 40 min, replicating the XRD measurement parameters. For the *in-situ* measurements, the 633 nm laser line was used with Zeiss LD EC Epiplan-Neofluar Dic x50 (NA 0.55) lens, 600 lines/mm grating, 30 s integration, 3 accumulations, with 4 measurement points taken, and the resulting spectra were averaged. The final step of Raman spectroscopy measurements was *post-mortem* Raman imaging with the use of two excitation laser lines (633 and 488 nm), Zeiss EC Epiplan-Neofluar Dic x100 (NA 0.9) lens, 600 lines/mm grating, 0.25 s (633 nm) or 2 s (488 nm) integration time. The size of the maps was adjusted to the size of the region of inters on the surface of the samples (20x15 µm) and on the cross-section (20x9 µm), whereas the measurement lateral step was fixed to 0.5 µm. Additional measurements were performed on the fully-crystallised coating after long-term (2 weeks) thermal treatment.

### *3. Results and discussion*
#### *3.1. High-temperature X-ray diffraction*

The investigation of high-temperature diffraction was conducted within a dedicated chamber featuring a graphite dome. Consequently, characteristic peaks of the graphite material were detected and labelled on diffractograms, alongside substrate signals marked as steel (Fig. 1a) [21]. To facilitate a clear interpretation of the obtained results, fragments containing peaks of interest were highlighted while peaks that were characteristic for graphite and steel were excluded (Fig. 1b, c, d) [22]. The recorded data indicated that the material was X-ray amorphous up to temperatures of at least 650 °C. At 750 °C, the first new peaks emerged, indicating the initiation of the crystallisation of the coating. Signals at approximately 37°, 39°, 46°, and 67° were identified as γ-$Al_2O_3$ [23, 24]. They correspond to the (311), (222), (400), and (440) *hkl* positions of the cubic γ-$Al_2O_3$ in the *Fd-3m* space group [25, 26].

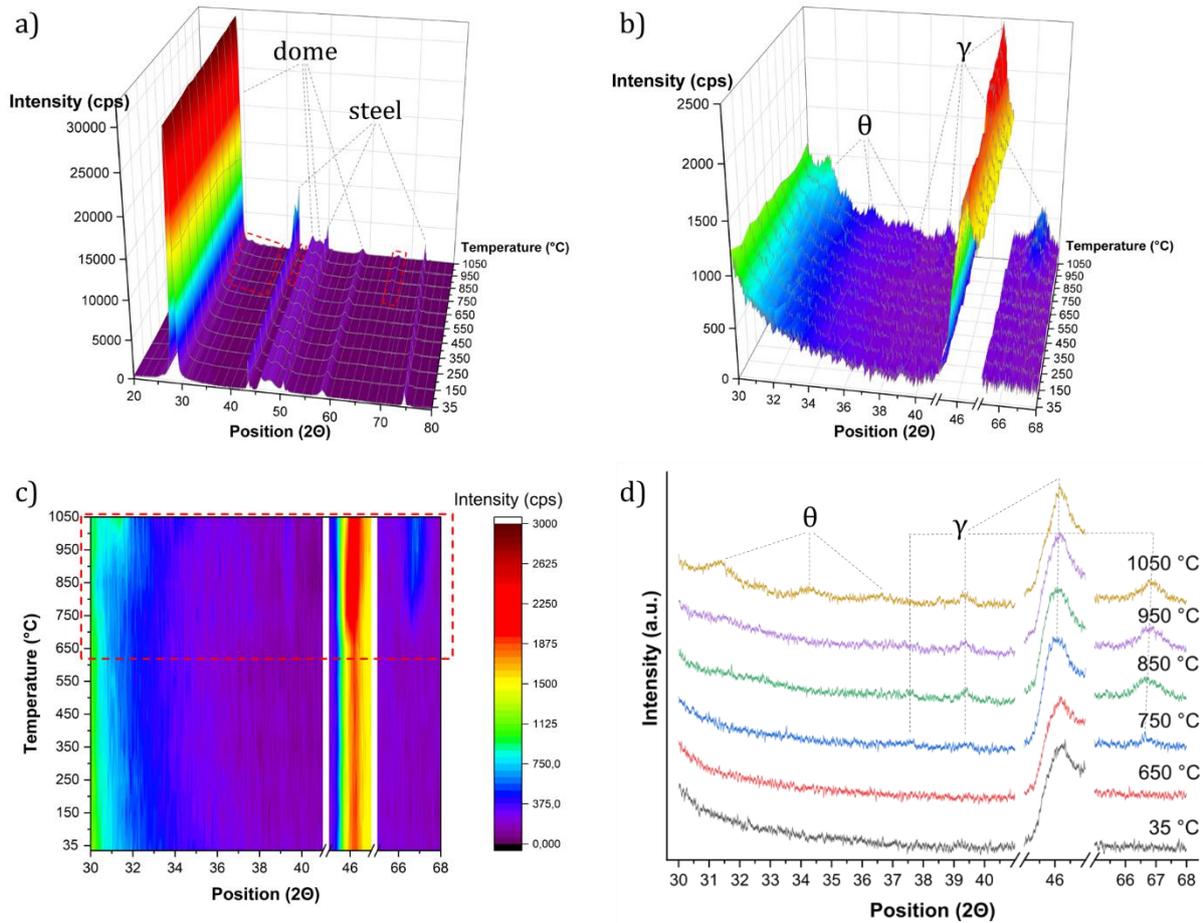

*Figure 1. HT-XRD analysis results: recorded structural evolution of the alumina coating deposited on the steel substrate, with marked red dashed rectangle regions of interest (a) depicted further in the form of 3D diagram (b) and the top view (c), diffractograms from crucial temperatures with assignments to the identified phases (d)*

At the highest temperature examined (1050 °C), the monoclinic *C2/m* θ-$Al_2O_3$ phase was observed, as evidenced by three characteristic peaks at 31.4°, 34.2°, and 36.6° corresponding to (400), (204), and (221) planes [26]. It is worth noting that the peak at 46° partially overlapped with a strong signal from the graphite dome. However, its shape evolved with temperature, indicating the presence of a new peaks. No intermediate or subsequent phase transformations were observed via XRD, although the formation of the α-$Al_2O_3$ phase was expected at temperatures above 900 °C [26]. Nonetheless, its presence was not detected under the applied experimental conditions, suggesting that the analysed alumina coating exhibited satisfactory thermal stability. However, further experiments indicated that this material required significantly longer durations at subsequent temperatures to undergo phase transformations, presumably several hours.

*3.2. Optimization of methodology of Raman spectroscopy measurements*

The WITec microspectroscope and TS1500 HT chamber were used to record high-temperature Raman spectra. *In-situ* analysis of alumina-based coatings is challenging due to the components of the HT chamber being manufactured form $Al_2O_3$-containing materials. The chamber features a sapphire transparent top closing window, which interferes with the results obtained from the analysed sample, even though it is not in the laser focal point, as shown in the preliminary study (Fig. 2). The obtained spectra (Fig. 2a) exhibited structural bands (located below 800 cm$^{-1}$) and emission bands (above 1300 cm$^{-1}$) characteristic of α-$Al_2O_3$ [13, 19, 27, 28].

The vibrational phonon modes of α-Al$_2$O$_3$, belonging to the D$_{3d}$ space group, are represented by the following modes: 2A$_{1g}$+2A$_{1u}$+3A$_{2g}$+2A$_{2u}$+5E$_g$+4E$_u$. According to the selection rules, 2A$_{1g}$+5E$_g$ modes are Raman active, resulting in seven visible phonon bands on the spectrum (Fig. 2b) [17]. The emission spectrum (Fig. 2c) may be recalculated to wavelength for facile comparison to the literature data and spectra obtained with other techniques such as photoluminescence spectroscopy [29]. The described effect of the mirage α-alumina spectrum is strictly connected with the window's material and might be also indirectly related with mirror-like alumina coatings surface. Therefore, the sapphire window was replaced with analogous silica window.

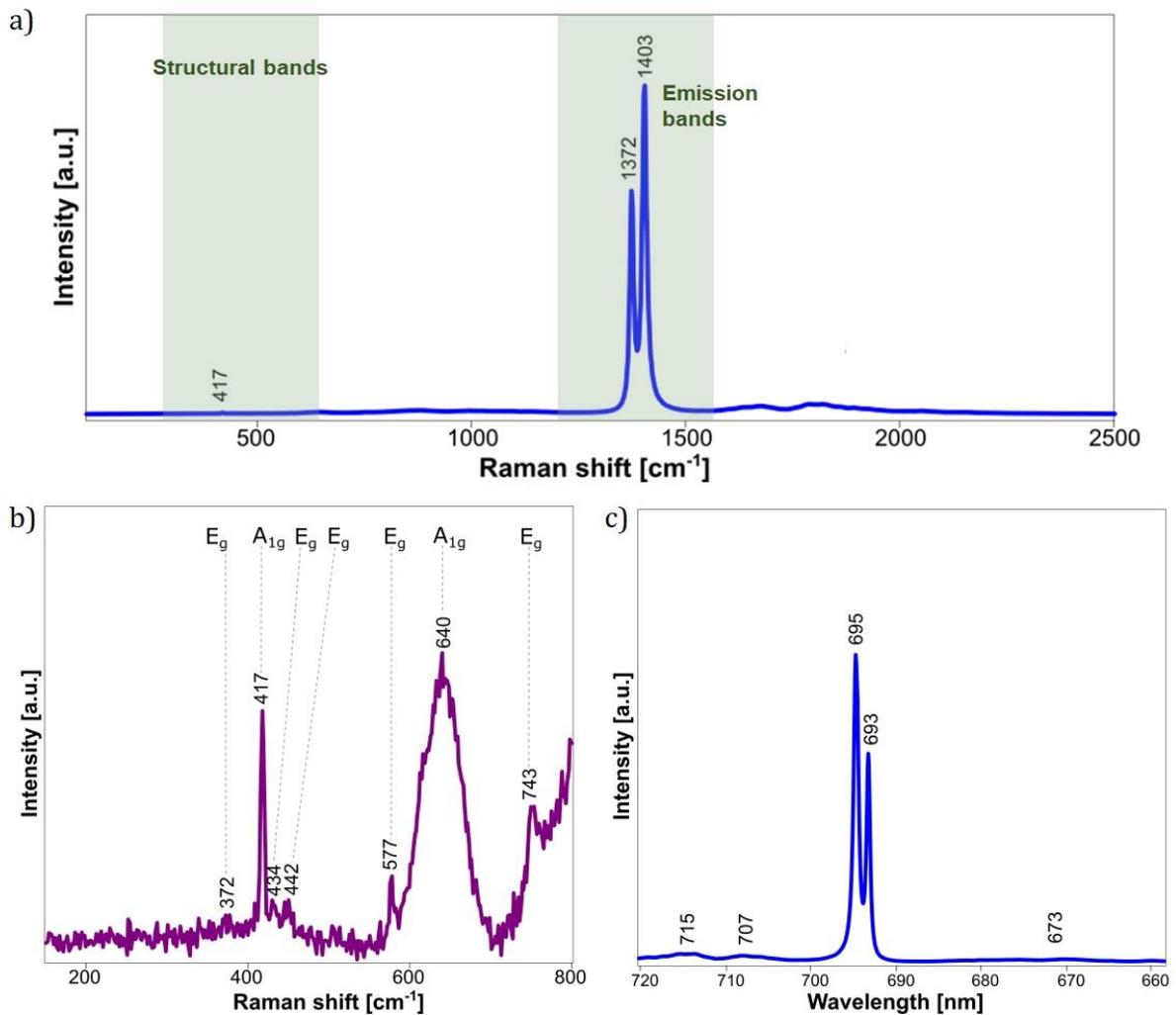

*Figure 2. Room-temperature Raman spectrum registered with Linkam TS1500, representing the sapphire window (not the coating) (a), spectral region with structural bands (b); emission spectrum with recalculated units (c).*

With this obstruction eliminated, another factor that might hinder the results of the measurements is the chamber's lining manufactured from corundum. At high temperatures, the high non-characteristic emission of the lining obscure the signal with the origin at the sample. The registered spectra possess elevated background at higher wavenumbers, as evidenced in Figures 3. No bands were observed which might indicate the presence of the alumina polymorph with Raman-active modes, proving efficiency of changing the top chamber window. Moreover, this finding is consistent with the XRD patterns, which reveal an amorphous coating. The initial high-temperature measurement methodology involved heating the sample at a rate of 5 °C/min and maintaining it at the set temperature for a short period (5 min). The complete evolution of the spectra (Fig. 3) shows subtle fluorescence at low temperatures and significant thermal effects above 800 °C with a raised background signal at higher Raman shift value.

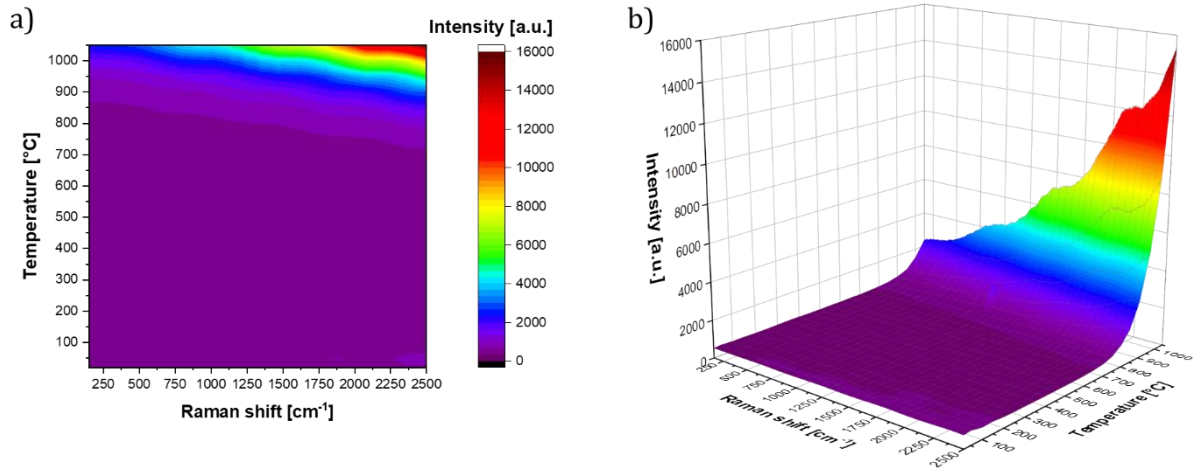

*Figure 3. In-situ high-temperature Raman spectroscopy results with short maintenance at desired temperatures: full thermal evolution - topographic colour graphs 2D (a) and 3D (b).*

The surface of the sample after the short-term high-temperature treatment was examined using Raman imaging (Fig. 4). The emission spectrum, measured with a red laser (633 nm) reviled the presence of α-alumina and trace amounts of γ-alumina in a non-uniform distribution (Fig. 4c) [13, 27, 28]. In areas with a lower presence of crystalized alumina polymorphs on the coating's surface, the products of substrate oxidation were observed in the form of magnetite (Fig. 4b) [21, 30]. However, the exact shape and low intensity of the spectrum of magnetite, coupled with the fact that it was discovered during surface imaging rather than point measurements, indicate very low quantity of it. Moreover, no other common oxidation product was found. That might suggest that the origin of the iron oxide is in the PLD alumina deposition rather than the applied thermal treatment.

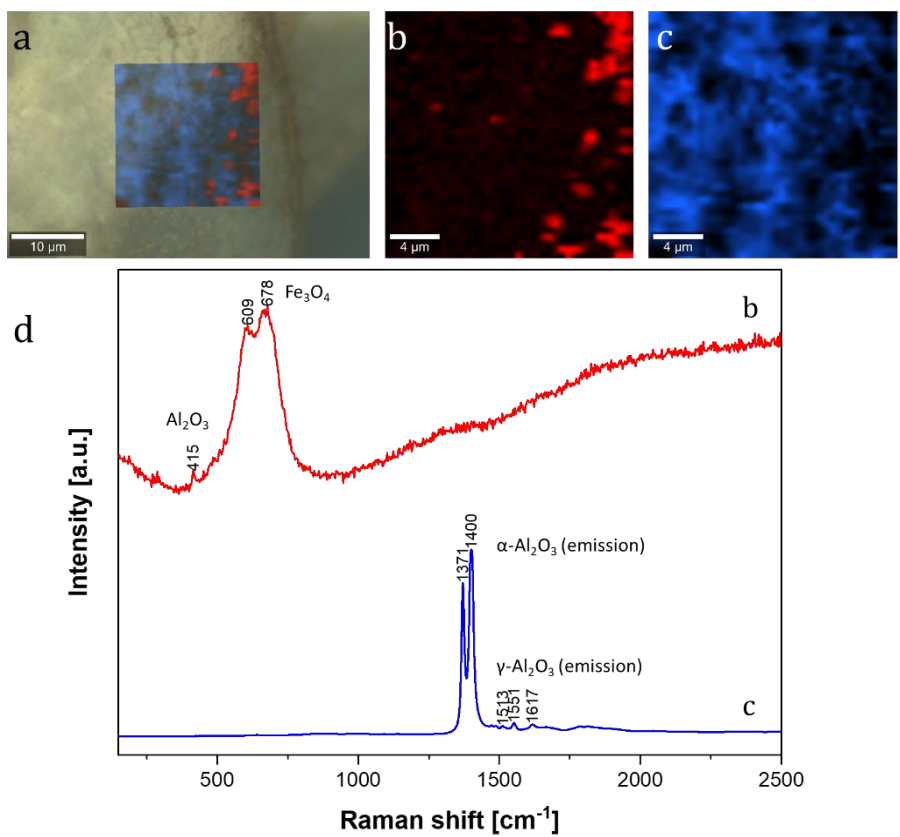

*Figure 4. Raman imaging of the coating after short-term HT-treatment: overlap of distribution maps and the microscopic image (a), distribution of substrate oxidation products obtained with 488 nm laser (b), distribution of α-alumina based on the emission spectrum excited with a 633 nm laser (c), corresponding Raman spectra (d).*

### 3.3. High-temperature Raman spectroscopy with medium-length high-temperature treatment

For the second stage of the *in-situ* HT-Raman spectroscopy measurements, we replicated the conditions used for HT-XRD with a maintenance time at each of the set temperatures prolonged to 40-minute. The critical analysis range covered temperatures close to room temperature and ranged between 550 and 1050 °C (Fig. 5). As previously mentioned, no visible structural bands were present prior to the HT-measurements. Under heating, the elevated background again hinders the possibility to clearly observe changes in the structure of the coating occurring under elevated temperatures. However, the change in the regularity of the scope of the spectra between 1350-1420 cm$^{-1}$ at highest temperature (Fig. 5b) may suggest crystallization of α-alumina [13, 19,27, 28]. Nevertheless, the post-mortem analysis of the sample was necessary to characterize its structural evolution in detail.

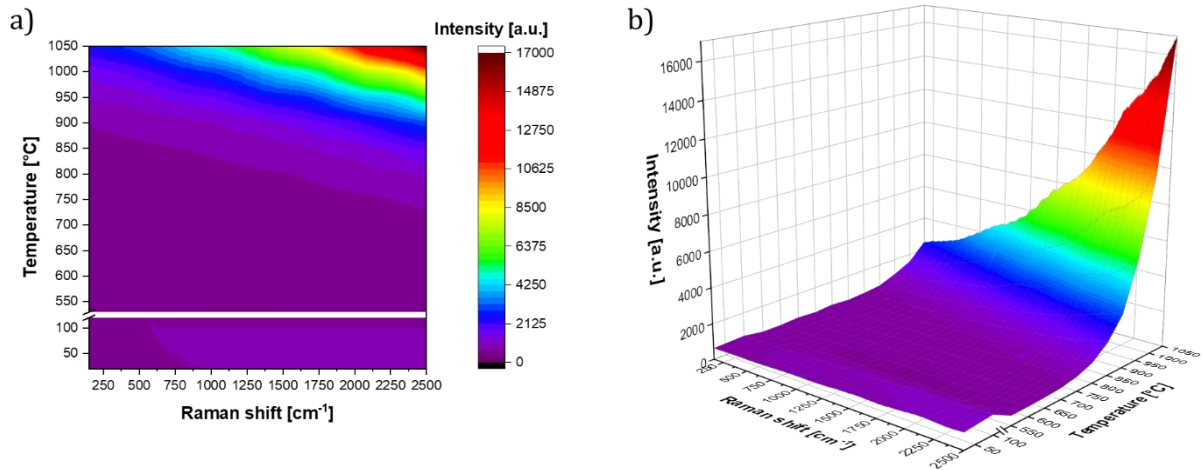

*Figure 5. In-situ high-temperature Raman spectroscopy results for elongated to 40 min maintenance at desired temperatures: 2D (a) and 3D (b) plots.*

After the thermal treatment, the coating remained with no visible exfoliation to the naked eye. Therefore, it was possible to prepare a cross-section of the coating and base imaging on such a prepared sample (Fig. 6). During the analysis, it was reviled that the amorphous coating crystallised rather uniformly to α-alumina with minor addition of γ-$Al_2O_3$ (Fig. 6e) [27, 28]. Bands with the origin in amorphous carbon (Fig. 6d) where observed in several spots in the internal part of the coating neighbouring the localization of $Fe_3O_4$ (Fig. 6b) as a steel substrate oxidation product, drawing strong correlation [21, 30]. Other oxidation products were observed in more outer part of the coating (Fig. 6c). The scale contained chromium oxide, (Fe/Cr)$MnO_4$ spinel, and hematite [21, 30, 31]. All of the oxidation products were accumulated along straight line perpendicular to the substrate. It suggests that in this particular place, a specific oxidation path occurred, which might be connected with a cracking or localized weakening of the integrity of the coating. Other areas of the cross-section were devoid of oxidation products, showing no signs of substrate compounds (e.g., chromium) diffusing through the coating.

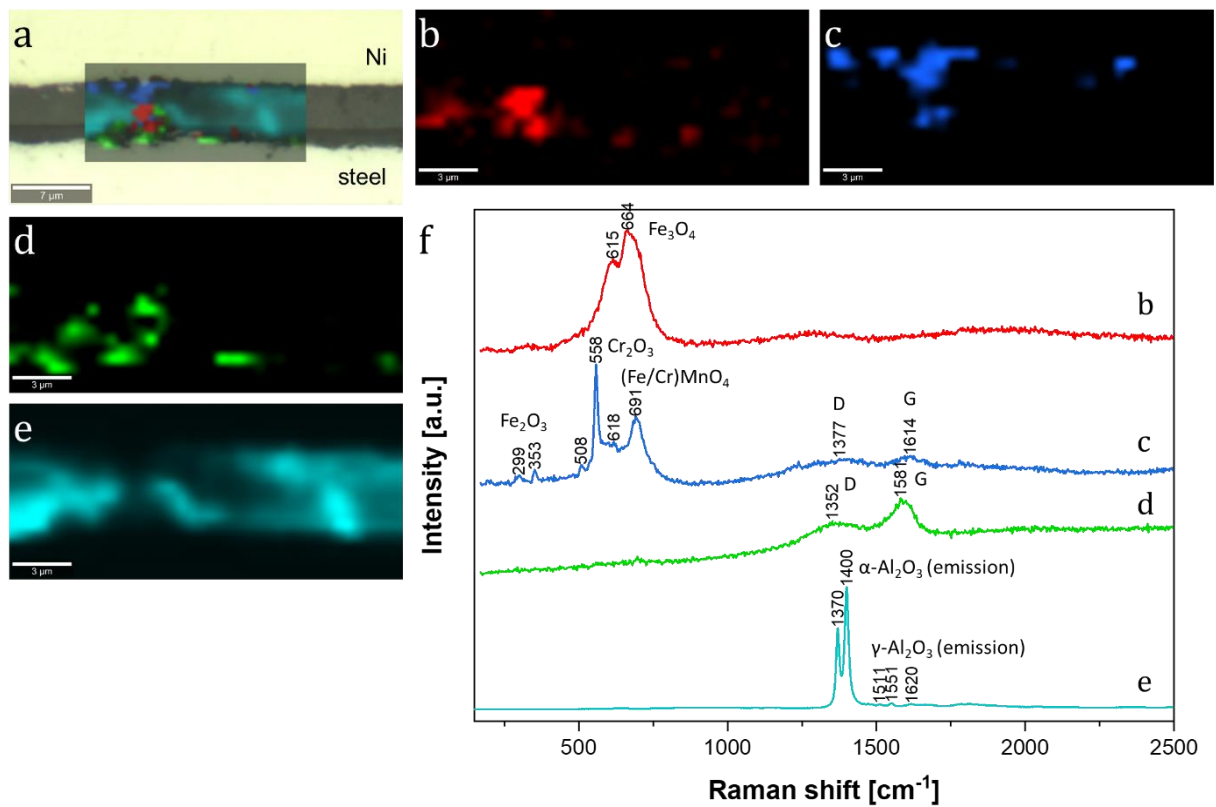

*Figure 6. Raman imaging of the cross-section of the coating after the medium-length thermal treatment: overlap of distribution maps and the microscopic image (a), distribution of substrate oxidation products (b, c), and amorphous carbon (d) obtained with 488 nm laser, distribution of α-alumina based on the emission spectrum excited with a 633 nm laser (e), corresponding Raman spectra (f).*

As a reference, strongly oxidized sample coating after an extended thermal treatment was examined. The coating had signs of exfoliation, and the scale was relatively fragile, making it impossible to prepare a cross-section. Observations were conducted on the surface (Fig. 7). The presence of chromium oxide (Fig. 7f) and spinel (Fig. 7c) were identified relatively uniformly distributed [19, 21, 30, 31]. The character of the spinel was analogous to the one found before (Fig. 6c). Additionally, crystalized α-alumina and a minor quantity of γ-alumina were present on the surface (Fig. 7d) [13, 27, 28]. In contrast to the above-described samples after shorter thermal treatment, the presence of θ-alumina was detected (Fig. 7e), localised in the central part of the analysed region [16].

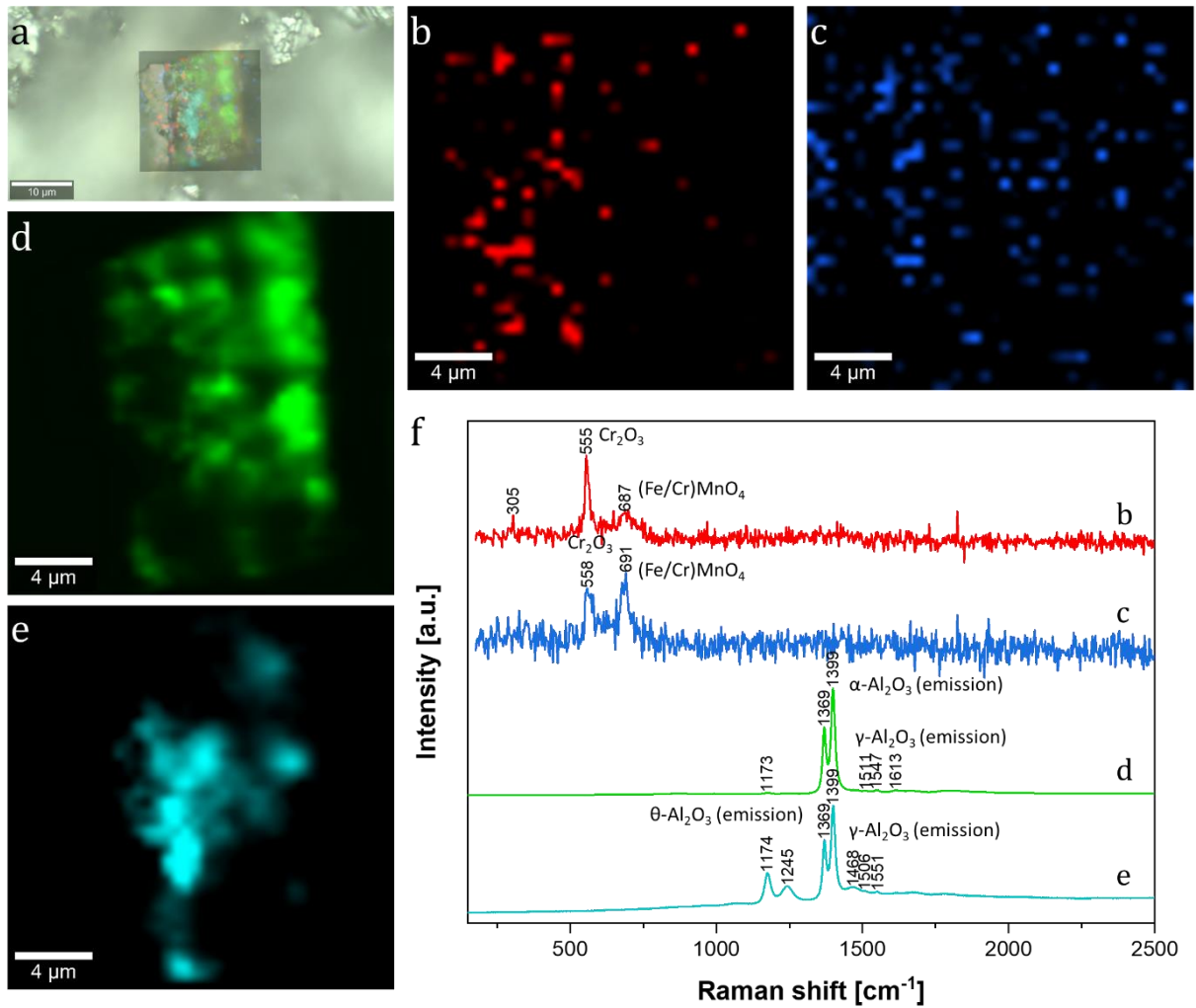

*Figure 7. Raman imaging of the surface of the coating after the most extended thermal treatment: overlap of distribution maps and the microscopic image (a), distribution of substrate oxidation products (b, c) obtained with a 488 nm laser, distribution of α-, θ- and γ-alumina based on the emission spectrum excited with a 633 nm laser (d, e), corresponding Raman spectra (f).*

## 4. Conclusions

The study investigated the thermal stability and structural integrity of amorphous alumina coatings for potential application in Generation IV nuclear reactors, particularly in Lead-cooled Fast Reactors (LFR). A 5 μm alumina coating, pulsed laser deposited on a 316L stainless steel substrate, was analysed *in-situ* and *post-mortem* using high-temperature Raman spectroscopy and X-ray diffraction (XRD). Raman imaging was also employed to examine the surface and cross-section of the coating after thermal treatments of varying lengths.

The study revealed the initial amorphous character of the coating. The presence of α-alumina with additional polymorphs, γ- and θ-$Al_2O_3$, was detected only after prolonged high-temperature exposure. HT-XRD analysis showed the coating to be X-ray amorphous up to 650 °C, with crystallization initiation at 750 °C. The presence of γ-$Al_2O_3$ and θ-$Al_2O_3$ phases was observed at higher temperatures. Structural changes of alumina were disclosed with HT-Raman spectroscopy and Raman imaging. No rapid transformations were observed. *Post-mortem* examination indicated the presence of α -$Al_2O_3$ and γ -$Al_2O_3$ with traces of θ-$Al_2O_3$. The small quantity of the later, compared to the *in-situ* HT-XRD, indicates its phase transition into α-$Al_2O_3$. The presence of oxidation products was correlated with long high-temperature exposure and occurred locally.

Oxidation signs were detected at temperatures exceeding LFR reactor working conditions. Moreover, the coating exhibited satisfactory thermal stability up to 1050 °C.

The alumina coating demonstrated great potential as protective layer in LFR nuclear reactors, showing good thermal stability and resistance to oxidation. The combination of high-temperature Raman spectroscopy and XRD provided comprehensive insights into the structural evolution of the coating under elevated temperatures. For future work, *in-situ* or *ex-situ* investigation of the PLD alumina deposition process is warranted to examine formation of the interface between the coating and the steel substrate. Furthermore, longer durations at subsequent temperatures and different thicknesses of the $Al_2O_3$ layer may be explored for a detailed understanding of alumina phase transformations.

The presented study contributes valuable data on the behaviour of alumina coatings under high-temperature conditions, essential for evaluating their potential application in advanced nuclear reactor designs. Insights gained from *in-situ* and *post-mortem* analyses provide a basis for understanding the longevity and performance of the coatings in the challenging environment of LFR reactors. Moreover, conscious optimization of the methodology was refined by acknowledging limitations and adjusting the Raman spectroscopy conditions, including changing the chamber window to silica to eliminate interference.

*Acknowledgement*

The research leading to these results was carried out in the frame of EERA Joint Programme on Nuclear Materials and is partly funded by the European Commission Horizon 2020 Framework Programme under grant agreement No. 755269 (GEMMA project) and grant agreement No. 101061241 (INNUMAT).

The authors (MG, AZ, RD, ŁK) were supported partially by the European Union Horizon 2020 research and innovation program under Grant Agreement No. 857470 and from the European Regional Development Fund under the program of the Foundation for Polish Science International Research Agenda PLUS, grant No. MAB PLUS/2018/8, and the initiative of the Ministry of Science and Higher Education 'Support for the activities of Centers of Excellence established in Poland under the Horizon 2020 program' under agreement No. MEiN/2023/DIR/3795.